\documentclass[twocolumn]{revtex4}
\usepackage{dcolumn}
\usepackage{graphicx}
\usepackage{amsmath}
\begin{document}

%
%
%
%
%

\title{Dynamics of viscoelastic membranes}
\author{Alex J.~Levine}
\affiliation{Department of Chemical Engineering, University of
California, Santa Barbara. Santa Barbara CA 93106}

\author{F.C. MacKintosh}
\affiliation{Division of Physics \& Astronomy, Vrije Universiteit
1081 HV Amsterdam, The Netherlands} \affiliation{Department of
Physics, University of Michigan, Ann Arbor, MI 48109-1120}

\date{\today}
\begin{abstract}
We determine both the in-plane and out-of-plane dynamics of
viscoelastic membranes separating two viscous fluids in order to
understand microrheological studies of such membranes. We
demonstrate the general viscoelastic signatures in the dynamics
of shear, bending, and compression modes. We also find a screening
of the otherwise two-dimensional character of the response to
point forces due to the presence of solvent. Finally, we show
that there is a linear, hydrodynamic coupling between the
in-plane compression modes of the membrane and the out-of-plane
bending modes in the case where the membrane separates two
different fluids or environments.
\end{abstract}

\maketitle

\section{Introduction}

Membranes and biopolymers form many of the most basic structures
of plant and animal cells. These fundamental building blocks
frequently occur together in complex structures. In animal cells,
for instance, the outer membrane is often strongly associated with
a network of filamentous actin, one of the most prevalent proteins
in the cell. This actin cortex is viscoelastic and it contributes
significantly to the response of whole cells to external stress.
Many prior physical studies, both theoretical as well as
experimental, have concerned the structure and dynamics of simple
membranes\cite{Brochard:75} and interfaces\cite{Chou:95}. Much
less is known about complexes of membranes with biopolymers.
Recent experiments have demonstrated the ability to both construct
and probe in vitro models consisting of lipid membranes with
attached actin biopolymer\cite{Helfer:00}. By a microrheology
technique, the material properties of these micrometer-scale
viscoelastic films could be measured. Here, we calculate the
dynamics, and thereby the response, of viscoelastic membranes. We
demonstrate, among other things, the general signatures of
viscoelasticity in the dynamics of both shear and bend. We also
find a coupling of bend and compression modes when there are
different environments on the two sides of the membrane, as is the
case for the surface of a biological cell. These effects have
important implications for previous and ongoing microrheological
studies of both the model biopolymer--membrane complexes, as well
as real cells.

Microrheology\cite{Mason:95,Mason:97,MacKintosh:99} studies the
rheological properties of a material by the use of small probe
particles, which can either be actively manipulated by external
forces\cite{Sackmann:94}, or imaged while subject to thermal
fluctuations. It is possible, for instance, to extract the
viscoelastic moduli from observations of the fluctuating position
of a small (Brownian) particle embedded in the
medium\cite{Mason:95,Mason:97,Gittes:97,Schnurr:97,Crocker:00}.
The technique holds out great promise as a new biological probe
measuring the material properties of living
cells\cite{Sackmann:94}.  In effect, the judicious application of
this technique may permit the creation of a ``rheological
microscope,'' providing new insights into the regulation and
time--evolution of the mechanical properties of various
intracellular structures over the course of the cellular
lifecycle, or in response to various external stimuli.  Such
studies of the cyctoplasm are already
underway\cite{Kuo:00,Valentine:01}. One essential feature of the
cell is the cell membrane, an essentially two--dimensional lipid
bilayer incorporating a wide variety of dissolved proteins and
anchored to a cytoskeletal network. Microrheological studies of
artificial biopolymer--membrane complexes\cite{Helfer:00} have
been performed, and efforts are under way to extend such
techniques to the membranes of real cells\cite{Wirtz:01}. In order
to study the rheology of the cell membrane using microrheological
techniques, the previously--studied methods used to extract
rheological measurements from thermal fluctuations in
three--dimensional
samples\cite{Mason:97,Gittes:97,Schnurr:97,Crocker:00,Levine:00,Levine:01}
need to be extended to problem of a viscoelastic membrane coupled
to a viscous solvent.

In this paper we consider such an extension of these ideas, with
an eye toward not only cellular microrheology, but also the
investigation of a wide variety of systems in soft physics
wherein a viscoelastic membrane is coupled to a viscous fluid
(typically water).  Examples include emulsions, vesicles, and
Langmuir monolayers. We calculate the position response to a
force of a small rigid particle embedded in the complex, soft,
viscoelastic medium. Using the Fluctuation--Dissipation
theorem\cite{Lubensky-book} we can then compute the
autocorrelations of the probe particle's position -- a quantity
accessible in experiments. As in previous work in this
field\cite{Brochard:75,Gittes:97,Schnurr:97,Helfer:00,Levine:00,Levine:01},
careful attention must be paid to the full linear mode structure
of the system. The observed thermal fluctuations of the probe
particle are in response to all such modes that couple to the
particle position thus the full mode spectrum must be determined
in order to interpret microrheological data.  This is in
distinction to more traditional rheology, which is a linear
response measurement of the system to applied shear strain and
therefore measures the shear modulus directly. In particular, in
this membrane study we are required to calculate the hydrodynamic
flows (in the low Reynolds number limit) in the viscous fluid
above (superphase) and below (subphase) the membrane generated by
membrane distortions. The understanding of these flows is
essential to calculating the modes of the combined system and,
hence, the response function of the probe particle. This point
has been previously recognized by Brochard {\it et
al.\/}\cite{Brochard:75} in their calculation of the decay of
membrane bending perturbations when coupled to a viscous fluid,
and was also used in the interpretation of the recent experiments
on actin-coated membranes \cite{Helfer:00}.

Qualitatively, we find that the introduction of an viscous
subphase and/or superphase introduces a new length scale over
which shear waves in the two--dimensional membrane decay due to
the viscous damping of the three--dimensional fluid. The
appearance of a new length scale is, of course, not surprising
given that the ratio of a two-dimension membrane shear modulus
$\mu(\omega)$ and a three dimensional fluid shear modulus
$G(\omega) = - i \omega \eta$ yields a length.  In fact this
length has been commented upon previously in the context of
viscous films in contact with a viscous solvent
\cite{Lubensky:96,Stone:98}. In the case of membrane dynamics and
microrheology, it sets a probe particle-dependent, high frequency
limit, beyond which the dynamics are controlled entirely by the
solvent\cite{Helfer:00}. In general, this length sets a limiting
range of the two-dimensional strain or velocity field.

In addition to the role of membrane--liquid coupling in the shear
modes, we perform a similar analysis of bending modes of the
membrane.  We reproduce the Brochard {\it et al.\/} results for
the mode structure in the case that subphase and superphase fluids
have the same viscosity, and we assess the out-of-plane response
function in the manner outlined briefly above.  More remarkably,
we find that linear independence of the in--plane membrane
compression modes and the out--of--plane bending modes is
destroyed when the $\hat{z} \longrightarrow - \hat{z}$ symmetry of
the problem is broken by a membrane (lying in the $xy$ plane) that
separates two fluids of {\it different\/} viscosities. We go on to
calculate the response function in this more general case as well
as in the symmetric problem.  The asymmetric problem has clear
physical relevance to dynamics of cellular membranes as well as to
vesicles and microemulsions. Perhaps the most dramatic example of
the broken symmetry occurs in the study of Langmuir monolayer
dynamics where the subphase (typically aqueous) and the superphase
(air) have viscosities which differ by many orders of magnitude.

Finally we point out that it is well--known that in membranes
whose equilibrium shapes are not flat, there is a linear coupling
of bending modes to in--plane distortions due only to the
nontrivial geometry of the surface. We do not discuss this sort of
coupling in the present work and restrict the present analysis to
the dynamics of flat membranes leaving the role of curvature in
microrheology to a later paper.

The remainder of this paper is organized as follows: In section
\ref{modes} we discuss the coupled modes of the membrane viscous
fluid system.  We use this analysis of the mode structure in
section \ref{response-function} to determine response function of
a rigid particle embedded in the membrane in the symmetric case
where in--plane and out--of--plane motions decouple.  We then
break the subphase/superphase symmetry in section \ref{coupled}
and reanalyze the response function for the case where
out--of--plane forces can generate in--plane motion. Finally, in
section \ref{experimental} we compute the predicted position
autocorrelations of a particle embedded in a membrane whose
material properties are typical of the various classes of systems
mentioned above.  We pay particular attention to the system of an
actin--coated lipid bilayer that has been investigated by the
Strasbourg group\cite{Helfer:00}. We conclude and propose further
experimental and theoretical work in \ref{conclusions}.

\section{The modes of the system}
\label{modes}

We now determine the modes of the membrane coupled
hydrodynamically to the (typically aqueous) subphase and
superphase. The essential calculation of these involves solving
the equations of motion for the fluid above and below the membrane
given a certain deformation mode of that membrane.  We shall do
this for the three independent modes of membrane deformation.
These modes include two in-plane modes, shear and compression, as
well as one out-of-plane, bending mode of the membrane. We
consider first the modes of in-plane membrane deformation, {\it
i.e.\/} those that do not involve curvature of the membrane. As
long as the membrane is flat in equilibrium, these modes are all
linearly independent.  We shall see, however, that in systems
having fluids of different viscosities in the subphase and
superphase, there is a linear-order, hydrodynamic coupling between
in-plane compression mode and the out-of-plane membrane bending.
However, if the two fluids are identical restoring the $\hat{z}
\longrightarrow - \hat{z}$ symmetry of the problem, the three
modes mentioned above are all linearly independent.  We will study
the modes of the broken--symmetry problem in a later section.

\begin{figure}[bp]
\scalebox{0.40}{\includegraphics{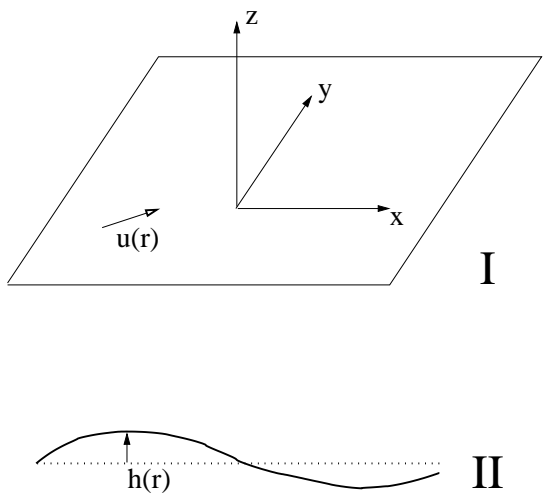}} \caption{The flat
membrane considered in this paper.  As shown in  (I) the membrane
lies in the $xy$ plane of out coordinate system.  The in--plane
displacement field ${\bf u}$ is defined in the plane of the
membrane.  The fluid superphase $z>0$ and subphase $z<0$ are not
shown. In (II) the edge--on view of the membrane shows the
vertical displacement $h$ of the membrane (solid line) from its
flat, equilibrium shape (dotted line). } \label{layout}
\end{figure}

The membrane lies in the $xy$ plane as shown in figure
\ref{layout}. The strain field ${\bf u}$ is a two--dimensional
vector lying in the plane of membrane.  We now need to determine
the fluid velocity field above the membrane ($z>0$) associated
with this shear wave. Working in the limit of zero Reynolds number
we solve the Stokes equation
\begin{equation}\label{Stokes}
  \eta \nabla^2 {\bf v} = {\bf \nabla} P,
\end{equation}
where the three dimensional vector ${\bf v}$ is the fluid velocity
field and $P$ is the hydrostatic pressure which enforces the fluid
incompressibility
\begin{equation}\label{incompress}
  \nabla \cdot {\bf v} = 0.
\end{equation}

These equations must be solved subject to the boundary conditions
\begin{eqnarray}
\label{boundary_conditions}
{\bf v}({\bf x}, z = 0, t) &=& \frac{\partial}{\partial t} {\bf
u}({\bf x}, t)\\ \label{boundary2}
 \lim_{z\longrightarrow \infty}
{\bf v}({\bf x}, z, t)&=& 0
\end{eqnarray}
reflecting the stick boundary conditions of the fluid at the
surface of the membrane and requirement that the fluid velocity
field go to zero at large distances from the membrane.

\subsection{Shear deformation}\label{shear-modes}

We choose membrane coordinates so that the shear wave propagates
in the $\hat{x}$ direction and the deformation is in the $\hat{y}$
direction. Thus the simple, in-plane shear deformation of membrane
is described by the strain field
\begin{equation}\label{shear}
  {\bf u}(y, t) = \hat{y} U_0 e^{i \left( q x - \omega t \right)}
\end{equation}

From the symmetry of the problem we look for a solution of
Eq.~{\ref{Stokes}) with boundary conditions given by
Eqs.~(\ref{shear},\ref{boundary_conditions},\ref{boundary2})of the
form
\begin{equation}\label{guess-shear}
{\bf v} = -i \omega U_0 \hat{y} f(z) e^{i ( q x - \omega t)},
\end{equation}
where $f(z)$ is an unknown function satisfying the conditions:
$f(0) = 1$, $\lim_{z \longrightarrow \infty} f(z) = 0$ so that the
ansatz, Eq.~(\ref{guess-shear}) satisfies the requisite boundary
conditions.  The Stokes equation demands that the vorticity of
fluid flow, $\nabla \times {\bf v}$, satisfies Laplace's equation
\begin{equation}\label{Laplace}
  \nabla^2 \left( \nabla \times {\bf v} \right) = 0.
\end{equation}

Using our ansatz, we find that unknown function $f(z)$ satisfies
the differential equation
\begin{equation}\label{diff}
  \frac{d^2 f}{dz^2} - q^2 f = 0.
\end{equation}
Along with the boundary conditions given above for $f(z)$, we find
a solution for the velocity field in the region above the membrane
($z>0$)
\begin{equation}\label{shear-solution}
  {\bf v}({\bf x},z,t) = -i \omega U_0 \hat{y} e^{i q x - \left|q\right| z} e^{-i
  \omega t}.
\end{equation}
The fluid velocity in the subphase ($z<0$) is similar with $z
\longrightarrow - z$. Returning to the Stokes equation we find
that there is no pressure gradient associated with this fluid
motion. We see that the shear flow induced in the three
dimensional viscous liquid phases decays over a distance
comparable to the wavelength of the in--plane shear mode. This
viscous damping by the surrounding fluid introduces a new decay
length for shear waves in the membrane even in the case that the
membrane were perfectly elastic, {\it i.e.\/} $\mu(\omega) =
\mu_0$, a real constant.  More on this later. We now turn the
other in-plane mode of the membrane, the longitudinal or
compression mode.

\subsection{Compression mode}\label{compression-mode}

We now apply a longitudinal compression wave in the membrane
having a strain field
\begin{equation}\label{compression}
  {\bf u}({\bf x}, t) = \hat{x} U_0 \cos (q x) e^{-i \omega t}.
\end{equation}
and determine the associated fluid velocity field in the
superphase ($z>0$).

From an examination of the fluid flow near the fluid/membrane
boundary, we note that the compression mode injects a sinusoidally
varying vorticity field that is directed along the $\hat{y}$--axis
and varying in the $\hat{x}$ direction.  Since the vorticity must
satisfy the Laplace equation [Eq.~(\ref{Laplace})] and since the
boundary conditions require it to vary sinusoidally in the
$\hat{x}$ direction, the fluid vorticity must decay exponentially
into the fluid, {\it i.e.\/} in the $\hat{z}$ direction.  Based on
these considerations, we expect that the vorticity ${\bf \Omega} =
\nabla \times {\bf v} $ takes the form
\begin{equation}\label{vorticity}
{\bf \Omega} = \tau^{-1} \hat{y} \cos (q x) e^{- \left| q \right|
z}
\end{equation}
in the superphase. The fluid velocity must vanish at large
distances from the membrane so we have chosen a decaying
exponential in the $\hat{z}$ direction in Eq.~(\ref{vorticity}).
The constant $\tau$, with dimensions of time is, as yet,
undetermined. It will be selected to enforce the stick boundary
conditions of the fluid at the membrane's surface. From the above
equation and incompressibility we find the differential equation
obeyed by the $z$--component of the fluid velocity field in the
superphase
\begin{equation}\label{Vz}
  \left( \partial_x^2 + \partial_z^2 \right) v_z = \frac{q}{\tau}
  e^{- |q | z} \sin (q x).
\end{equation}
Once again, we use the ansatz for $v_z$:  $ v_z(x,z) e^{i \omega
t} = f(z) \sin ( q x )$ and find a differential equation of for
$f(z)$ of the form
\begin{equation}\label{diff2}
  \frac{d^2 f}{dz^2} - q^2 f =\frac{q}{\tau}  e^{- |q | z}.
\end{equation}
The boundary conditions at the membrane and at infinity require
that $f(0) = \lim_{z \longrightarrow \infty} f(z) = 0$.

\begin{figure}[bp]
\scalebox{0.60}{\includegraphics{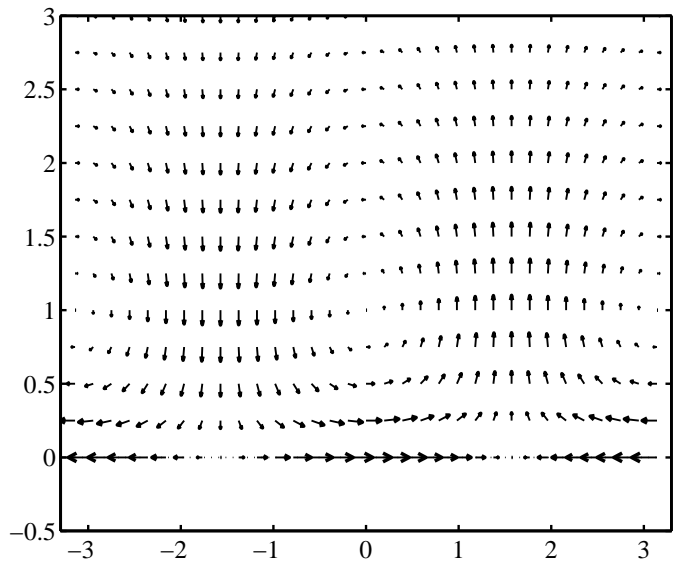}} \caption{The
membrane compression mode with associated flow field in the
superphase.  The membrane, seen edge-on as a dotted line undergoes
a compression wave of wavelength $2 \pi$.  The vector field
represents the fluid flow in the superphase and shows the
exponentially decaying vorticity.} \label{compression-figure}
\end{figure}

The homogeneous solution of the above boundary condition vanishes
upon the application of the boundary conditions on $f(z)$.  This
leaves only the particular solution to give the solution for $v_z$
in the superphase.  Integrating the fluid incompressibility
equation, Eq.~(\ref{incompress}),  then gives the accompanying
solution for $v_x$. We find the fluid velocity to be given by:
\begin{eqnarray}\label{vx-solution} v_z(x,z,t) &=& -i \omega U_0  e^{-i \omega t}
  \left[ 1 - |q| z \right] e^{- |q| z} \cos (q x)\\
  \label{vz-solution}
  v_z(x,z,t) &=& -i \omega U_0 q e^{-i \omega t} z e^{- |q| z} \sin
  (q x).
\end{eqnarray}
Returning to the Stokes equation we can calculate the pressure
field associated with the above flow.  It is interesting to note
that there is a sinusoidally varying pressure field at the surface
of the membrane.  We find that, for the compressional wave in the
membrane introduced above, the pressure at the upper surface of
the membrane ($z \longrightarrow 0^+$) takes the form
\begin{equation}\label{pressure-compression}
  P(x,t) = 2 i  \eta \omega q U_0 \sin (q x) e^{ - i \omega t}.
\end{equation}
It appears from the hydrodynamic flow field shown in
figure~\ref{compression-figure}, from and pressure field
calculated above in Eq.~(\ref{pressure-compression} that there
should be a membrane deformation along its normal accompanying the
membrane compression mode.  Of course, in the symmetric case where
the fluids in the superphase and subphase are identical, such a
normal deformation must vanish by symmetry. As soon as this
symmetry is broken, however, there should be a membrane height
fluctuation in response to the longitudinal modes of the membrane.
This is, in fact, incorrect.  A complete calculation of the $z z$
component of the fluid stress tensor at the surface of the
membrane shows that the pressure term is exactly canceled by the
viscous stress arising from the gradient of the upward fluid
velocity.   Nevertheless, such a {\it linear order}, purely
hydrodynamic coupling between membrane compression and bending
modes  exists and is  mediated by the surrounding fluid as we
shall see in our analysis of the bending modes.  There the flows
generated by the membrane bending generates a viscous shear
stress at the surface of the membrane which couples to the
in--plane compression mode.

This effect is unrelated to the coupling of shear and bending
modes in membranes with finite mean curvature\cite{Deutsch}. The
new linear coupling could be realized in a variety of experimental
situation. Among these are microemulsions, polymersomes, cell
bodies, and Langmuir monolayers. These systems can break the
symmetry of the problem even in the limit of flat membranes (or
membranes with a large radius of curvature leading to a
vanishingly small geometric coupling between shear and membrane
bending) by having fluids of differing viscosities on either of
the membrane.  We return to this point later; we point out only
the fundamental asymmetry of the coupling.  Bending modes generate
in--plane compression, but in--plane compression does {\it not}
produce membrane bending.

\subsection{Bending mode}\label{bending-mode}

In this section we recapitulate the results of Brochard {\it et
al.\/} on the effect of hydrodynamics on the dynamics of membrane
bending modes.  To do this we apply a sinusoidal height
fluctuation to the membrane of the form
\begin{equation}\label{height}
  h(x,t) = h_q e^{i ( q x - \omega t)}
\end{equation}
where $h(x,t)$ measures the displacement of the membrane surface
in the direction normal to its surface, {\it i.e.\/} the
$\hat{z}$-direction. See figure~\ref{layout}. Once again, using
the Stokes equation we calculate the fluid flow in the superphase
generated by such a membrane displacement. The vorticity generated
by the membrane motion must still satisfy Laplace's equation the
symmetry of the problem admits flows only in the $xz$ plane.  We
assume that the solution takes the form
\begin{equation}\label{bending-guess}
  {\bf v}(x,z,t) = \left[ v_x(z) \hat{x} + v_z(z) \hat{z} \right]
  e^{i (q x - \omega t)}.
\end{equation}

Using the Laplace equation for the vorticity and the
incompressibility condition we find two differential equations for
the unknown functions $v_x(z), v_z(z)$
\begin{eqnarray}\label{incompress-bending}
  i q v_x(z) + \frac{d v_z}{dz} &=& 0 \\
  \label{laplace-bending}
\frac{d^3 v_z}{dz^3} - i q \frac{d^2 v_z}{dz^2} - q^2 \frac{d
v_x}{dz} + i q^3 v_z &=& 0
\end{eqnarray}
Combining Eqs.~(\ref{incompress-bending},\ref{laplace-bending}) in
order to eliminate $v_x$ we arrive at a differential equation
governing $v_z$ alone
\begin{equation}\label{bending-vz}
\frac{d^4 v_z}{dz^4} - 2 q^2 \frac{d^2 v_z}{dz^2} +  q^4 v_z = 0,
\end{equation}
which has the solution:
\begin{equation}\label{vz-solution-bending}
  v_z(z) = C_1 e^{|q | z} + C_2 e^{ - |q| z} + C_3  z e^{|q| z} +
  C_4 z e^{-|q| z}.
\end{equation}
To satisfy the boundary condition at infinity we set $C_1 = C_3 =
0$.  After integrating Eq.~(\ref{incompress-bending}) to obtain a
solution for $v_x(z)$ and applying the stick boundary conditions
at the surface of the membrane, we return to
Eq.~(\ref{bending-guess}) to write the solution for the fluid flow
field in the superphase
\begin{eqnarray}\label{bending-answer-x}
v_x(x,z,t) &=&  \omega h_0 |q| z e^{-|q| z}
  e^{i(qx - \omega t)}\\
\label{bending-answer-y}
   v_z(x,z,t) &=& - i \omega h_0 \left[ 1 + |q| z \right] e^{-|q| z}
  e^{i(qx - \omega t)}\\
\end{eqnarray}
These flows are sketched in figure \ref{bending-figure}.  As
required, the component of the fluid velocity field tangent to the
membrane vanishes.

\begin{figure}[bp]
\scalebox{0.60}{\includegraphics{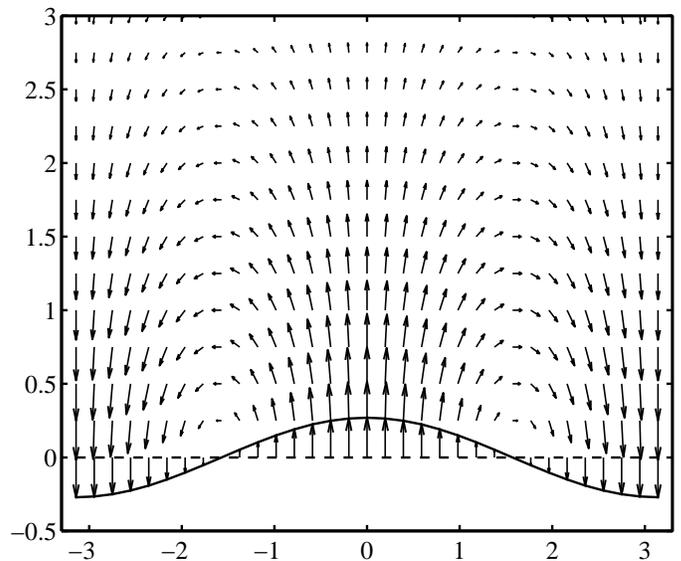}} \caption{The
membrane bending mode with associated flow field in the
superphase. The membrane, seen edge-on as a solid line undergoes a
bending wave of wavelength $2 \pi$. The undeformed membrane is
shown edge-on as a dotted line. The vector field represents the
fluid flow in the superphase. The component of the fluid velocity
field tangent to the membrane generates a shear stress that
couples to the in--plane compression mode.  This shear stress
changes sign across the membrane, however, only the flow in the
superphase is shown above.} \label{bending-figure}
\end{figure}

The pressure gradient associated with the above flow field can be
calculated directly from the Stokes equation.  We find, upon
setting the pressure to zero at infinity, that the pressure field
at the surface of the membrane takes the form
\begin{equation}\label{pressure-bending}
P(x,z=0,t) =  2 i \omega |q| \eta h_0 e^{i (q x - \omega t)}.
\end{equation}
We note in passing that fact that the amplitude of the pressure
oscillation is proportional to $q$ combined with the restoring
force on the bent membrane depending on wavevector as $q^4$ leads
to the well-known result that the decay rate of bending modes
increases as the third power of wavevector.

\section{The response function}\label{response-function}

We now turn the calculation of the response function.  The
calculation is similar in spirit to calculation of the response
function in three dimensions.  We wish to model the response to a
force of a particle embedded in the membrane at the origin. Given
the level of description of the membrane, {\it i.e.\/} a
continuum, single-component, two--dimensional viscoelastic medium,
there is no distinction between the application of a force to the
membrane itself over an area the size of the particle, and the
application of the same force to a rigid particle embedded in the
medium. Because of our reliance on a simple, coarse--grained
description of the membrane, we necessarily neglect any new
hydrodynamic modes of the membrane that may, in fact, be present
in the physical system. For example, if we were to consider a
two-component membrane composed of lipid and an elastic protein
network anchored in the lipid bilayer, there should be, in analogy
to the three--dimensional gel system, a ``free draining'' mode
associated with the diffusive relaxation of network density.
Nevertheless, we expect that such modes will be irrelevant at
higher frequencies where our single--component description of a
membrane should be valid for many systems of experimental
interest.  Details related to the more complex descriptions of
the membrane internal structure and their effect on
microrheological measurements we defer to a future publication.
In addition we do not consider the effect local perturbations of
the membrane structure (and consequently its viscoelastic
properties) due to the introduction of the probe particles. It has
been shown experimentally\cite{Crocker:00,Weeks:01} and
theoretically\cite{Levine:01a} in three dimensional systems that
such local perturbations can be important in the one particle
response function, but that interparticle response functions do
not depend on such effects to leading order.

We apply a force on the membrane that is localized at the origin
of the coordinate system and is sinusoidally varying in time. We
then compute the response of each Fourier mode of the membrane
deformation along with the associated fluid motion.  By
integrating over these motions we determine the motion of the
point at the origin of the membrane.  The response of each
Fourier mode in the viscoelastic membrane is determined from
force balance
\begin{equation}\label{force-balance}
  0 = \left[ \mu \partial^2 u_\alpha + (\mu + \lambda) \partial_\alpha
  \partial_\beta u_\beta \right]\delta_{\alpha i}^\perp -
  \delta_{i z}  \kappa \partial^4 h  + \sigma_{i z}^{\mbox{f}}|_{z=0} +
  f_i.
\end{equation}
In the above equation, the Greek indices run over the coordinates
in the plane of the membrane, $x,y$, while the Latin indices run
over all three coordinates.  The first term in square brackets on
the RHS of Eq.~(\ref{force-balance})  represents the in-plane
membrane viscoelastic force per area due to shear and compression
of the membrane.  The quantity $\delta_{\alpha i}^\perp$ projects
out the membrane coordinates. The membrane viscoelasticity is
described by two frequency--dependent Lam\'{e} constants
consistent with a isotropic continuum.  These terms should in fact
be expressed as integrals of over the strain history of membrane.
We will, however, suppress the frequency dependence of the
Lam\'{e} constants (and thus the viscoelasticity of the membrane)
until we rewrite the force--balance in the frequency domain.  The
force per area associated with the out-of-plane displacement of
the membrane, $h(x_\alpha,t)$, is given by the second term on the
RHS of the above equation, which represents the restoring force on
the membrane due to its bending rigidity, $\kappa$.  Once again,
in a viscoelastic membrane, we may assume that $\kappa$ is
frequency dependent, but we suppress this dependence for now.

Finally the last term on the RHS of Eq.~(\ref{force-balance}) is
the force per area exerted on the membrane by the external fluid
of the sub- and superphases.  The fluid stress tensor takes the
well--known for an incompressible, Newtonian fluid
\begin{equation}\label{stress-fluid}
\sigma_{i j}^{\mbox{f}}= \eta \left( \partial_i {\bf v}_j +
\partial_j {\bf v}_i \right) - P \delta_{ij}.
\end{equation}
The fluid velocity and hydrostatic pressure fields that accompany
any membrane motion has been determined in the previous section so
we may rewrite the fluid stress solely in terms of the membrane
displacement fields: ${\bf u}_\alpha$ and $h$.  Specializing to
the components the fluid stress tensor required in
Eq.~(\ref{force-balance}) and noting the stick boundary conditions
at the surface of the membrane we may simplify the form of the
full fluid stress tensor to
\begin{equation}
\label{stress-components} \sigma_{i z}^{\mbox{f}}= \eta \left.
\partial_z {\bf v}_i \right|_{z=0}  - P|_{z=0} \delta_{iz}.
\end{equation}

Because we are still considering symmetric case, the a force
acting in the plane of the membrane excites only the in-plane
deformation modes discussed above: compression and shear.
Similarly, the component of applied force acting along the
membrane normal only generates bending deformations.  Using this
decoupling of the in-plane and out-plane of plane modes we may
calculate the response to in-plane forces and forces along the
membrane normal ($\hat{z}$) separately. We begin with in--plane
forces

\subsection{In plane response}

We rewrite the force balance equation, Eq.~(\ref{force-balance})
specializing to the case of an in--plane force, $f_\alpha({\bf
x})$ and we Fourier transform in two--dimensional membrane
subspace. We arrive at
\begin{equation}
\mu q^2 u_\alpha({\bf q},\omega) + (\mu + \lambda) q_\alpha
q_\beta u_\beta ({\bf q},\omega) - \frac{\eta}{2}w_\alpha({\bf
q},\omega) = f_\alpha({\bf q},\omega),
\end{equation}
where we have defined $w_\alpha = \partial_z v_\alpha |_{z=0}$.
Projecting out the longitudinal and transverse parts of the above
equation and writing $u^{\rm L} = u_\alpha q_\alpha, u^{\rm
T}_\alpha = P^{\rm T}_{\alpha, \beta} u_\beta$ with $P^{\rm
T}_{\alpha, \beta} = \delta_{\alpha \beta} - \hat{q}_\alpha
\hat{q}_\beta$, we have
\begin{eqnarray}
\label{longitudinal} (2 \mu + \lambda) q^2 u^{\rm L} -
\frac{\eta}{2} w_\alpha \hat{q}_\alpha &=& f(q)_\alpha
\hat{q}_\alpha\\
\label{transverse}\mu  q^2 u^{\rm T}_\alpha -
\frac{\eta}{2} P^{\rm T}_{\alpha, \beta} w_\beta  &=&  P^{\rm
T}_{\alpha, \beta} f(q)_\beta
\end{eqnarray}

Using the results of section \ref{modes}, in which we have
computed the fluid flows associated with the longitudinal and
transverse modes of the membrane, we can determine the form of
${\bf w}$ in terms of the ${\bf u}$. In this way we can write
Eqs.~(\ref{longitudinal}),(\ref{transverse}) solely in terms of
${\bf u}$ and thus solve for the membrane displacement in terms of
the externally applied force, ${\bf f}$.  From the previous
section one finds:
\begin{eqnarray}
\label{longitudinal-flow} w_\alpha \hat{q}_\alpha &=& - i \omega
u^{\rm L}({\bf q}) \\ \label{transverse-q} P^{\rm T}_{\alpha,
\beta} w_\beta&=&  i \omega u^{\rm T}_{\alpha}({\bf q}) |{\bf q}|.
\end{eqnarray}

Combining the above equations with the force balance equations,
Eqs.~(\ref{longitudinal}),(\ref{transverse}) we solve for ${\bf
u}({\bf q}) = \hat{q} \hat{q} u^{\rm L}({\bf q}) +
(\tensor{\delta} - \hat{q} \hat{q} ) \cdot u^{\rm T}({\bf q})$. We
now integrate over the modes of the system to determine the
displacement of the point at the origin in response to the applied
force at that point.  The rigidity of the tracer particle at the
origin is accounted for by cutting off the wavevector integral at
the inverse radius of the embedded particle, $|q_{\rm max}| \sim
1/a$ or equivalently
\begin{equation}
\label{cutoff} f_\alpha({\bf q}, \omega) = {\bf F_0}_\alpha e^{-i
\omega t} \Theta ( q_{\rm max} - |{\bf q}|),
\end{equation}
where ${\bf F_0}$ is a constant vector, which for definiteness we
take to be in the $\hat{x}$ direction. We choose an order one
numerical prefactor in this relation between $q_{\rm max}$ and
$1/a$ so that the response function reproduces the standard stokes
drag on a spherical particle in the limit that the membrane
elasticity vanishes.  Having done this we find that the position
response  of the embedded tracer sphere (which is along the
$\hat{x}$ by symmetry) takes the form
\begin{eqnarray}
\label{integral-in-plane} u_x({\bf x}, \omega)  &=& F_0 \int
\frac{d^2 q}{(2 \pi)^2} \left\{ \frac{ \cos^2 \phi }{(2 \mu +
\lambda) q^2 - \frac{i \omega \eta}{2} |{\bf q}|} \right. \\
\nonumber
 & & + \left.
\frac{ 1 - \cos^2 \phi }{ \mu  q^2 - \frac{i \omega \eta}{2} |{\bf
q}|} \right\}
\end{eqnarray}
where the angle $\phi$ is defined by $\hat{r} \cdot \hat{x} = \cos
\phi$.  Clearly the integral is the response function that we seek
and via the Fluctuation--Dissipation theorem, it contains the
information necessary to determined the experimentally measured
power spectrum.  Performing the integrals we arrive at the final
form of the in--plane response function.  Since it is diagonal in
the in--plane indices we suppress them and write:
\begin{eqnarray}
\label{answer-in-plane} \alpha_{||}(\omega) &=& \frac{1}{4 \pi
\mu} \left[ \log \left( 1 + i  \frac{2 \mu}{3 \omega \eta a}
\right) \right.
\\
\nonumber
  & & \left. +
\frac{\mu}{2 \mu + \lambda} \log \left( 1 + i  \frac{2 (2 \mu +
\lambda) }{3 \omega \eta a} \right) \right]
\end{eqnarray}
We having chosen the wavevector cutoff here to be $q_{\rm max} =
2/(3 \pi a)$ so that in the limit of no membrane elasticity: $\mu,
\lambda \longrightarrow 0$ with $\mu/\lambda$ finite, the response
function reduces to the standard Stokes drag result from
low-Reynolds number hydrodynamics.

\subsection{Out--plane--response}

We now analyze the out--of--plane motion in a similar way by first
introducing a Fourier component of the bending deformation of the
membrane of the form of Eq.~(\ref{height}).  Using our
hydrodynamic results of section \ref{bending-mode} in combination
with Eq.~(\ref{force-balance}) and computing the hydrodynamically
induced stress using Eq.~(\ref{stress-components}) we determine
the amplitude of a bending mode of the membrane in response to an
arbitrary force. We then specialize to the case of a point force
and, integrating over the available bending modes of the system,
we determine the out--of--plane response of the point at the
origin to a force on it directed along the membrane normal.

We make use of the fluid flow field associated with the bending
deformation of the membrane as discussed in section
\ref{bending-mode}.  From this solution and
Eq.~(\ref{stress-components}) we find that at the surface of the
membrane there is a non-vanishing component of the fluid stress
tensor which takes the form
\begin{equation}
\label{bending-fluid-stress}
 \sigma^{\rm f}_{zz} = 2 i \omega
h_q|q| e^{i (q x - \omega t)}.
\end{equation}
It should be pointed out that although there is a nonvanishing
fluid stress component, $\sigma_{xz}^{\rm f}$, this stress changes
sign across the membrane and exactly cancels in the symmetric
case.  We return to this point later in the paper when we examine
the asymmetric problem in detail.  Furthermore, we point out that
if we were to consider a membrane as having finite thickness and
thus being able to support internal strains in which the
deformation gradient normal to the membrane's surface, {\it i.e.}
a membrane that can support a stress strain of the form $u_{xz}$,
then the bending mode would hydrodynamically couple at linear
order to such internal deformations of the membrane.  We do not
pursue this point in the current paper. For our present purposes,
it is enough to  note that a force normal to the plane of the
undeformed membrane couples only to the bending modes in the
symmetric case.

From the $z$ component of the force balance equation
[Eq.~(\ref{force-balance})] we find that a force in the $z$
direction with a  sinusoidal dependence on $x_\alpha$ and $t$
generates a sinusoidal membrane bending deformation with amplitude
\begin{equation}
\label{bending-amplitude} h(q) = \frac{f_z(q)}{\kappa q^4 - 2 i
\omega \eta |q| }.
\end{equation}
Integrating over such forces in q-space, determine the response to
a point force at the origin using the wavevector cutoff introduced
in Eq.~(\ref{cutoff})we find the out--of--plane response function
\begin{equation}
\label{out-of-plane-alpha} \alpha_z(\omega) = \frac{9 \pi a^2}{2
\kappa} \int_0^1 d p \frac{1}{p^3 - i \delta}
\end{equation}
where the dimensionless parameter $\delta = 27 \pi^2 a^3 \omega
\eta/(4 \kappa)$ is the ratio of the viscous stress $\omega
\eta/a$ to the membrane bending stress $\kappa/a^4$ at the length
scale of the probe particle.  We have now completed our discussion
of the response function of a bead embedded in the membrane for
the symmetric case.  In this case the complete motion of the
particle, ${\bf x}(\omega)$,  in the membrane is a linear
combination of the in--plane motion due to in--plane components of
the applied force and out--of--plane motion due to the component
of the externally applied force normal to the membrane. Thus the
most general solution of the mechanical problem to linear order in
membrane displacements takes the form
\begin{equation}
\label{final-result-symmetric} x_i(\omega) = \alpha_{||}(\omega)
(\delta_{ij} - \hat{z}_i
 \hat{z}_j ) f_j(\omega) +  \alpha(\omega)_z f_z(\omega)
 \end{equation}
 where the response functions $\alpha_\perp(\omega)$ and
 $\alpha_z(\omega)$ are given by Eqs.~(\ref{answer-in-plane}) and
 (\ref{out-of-plane-alpha}) respectively and ${\bf f}(\omega)$ is
 the externally applied force responsible for the motion.  We will
 later use the Fluctuation--Dissipation theorem to explicitly
 compute the implications of this result for the experimentally
 observed position fluctuations of the probe particle.  First, we
 turn to the case of the asymmetric problem to examine the coupling
 of in--plane membrane compression modes to the out--of--plane
 membrane bending modes and the implications of such coupling on
 the response function.

\section{Asymmetric system}\label{coupled}

Here we examine the system in which the membrane separates fluids
of differing viscosities.  In order to study membrane dynamics in
this asymmetric case we return to the question of the hydrodynamic
stresses on the membrane with a particular emphasis on the
coupling of compression and bending (out of plane motion) modes
mediated by the hydrodynamic coupling through the
subphase/superphase.  From our previous calculation of the fluid
velocity fields associated with the various deformation modes of
the membrane, we determine the viscous shear stress on the
membrane by demanding the continuity of the shear stress
$\sigma_{\alpha z}, \alpha= x,y$ at the membrane--fluid boundary.
In the fluid we note that this component of the stress tensor
takes the form
\begin{equation}\label{fluid-stress}
  \sigma^{\rm f}_{\alpha z} = \eta \partial_z v_\alpha.
\end{equation}
We note that $\partial_\alpha v_z = 0, \alpha = x,y$  due to the
fluid boundary conditions on the membrane.  Due to the linearity
of the hydrodynamics, we can combine the solutions of the flow
fields from the longitudinal (compression) and bending modes to
write the full shear stress associated with a linear combination
of those membrane deformations.  We may deal with the transverse
membrane modes separately since they do not couple to the bending
degrees of freedom as does the compression mode.  Finally, we note
that the shear flows on the opposite sides of the membrane
generated by the longitudinal in-plane modes contribute
additively, while the flows coming from the out-of-plane
deformation of the membrane produce shear stresses on either side
of the membrane that tend to cancel.  Both of these shear
stresses, however, are proportional to the viscosity of the fluids
on either side of the membrane; thus there will be a residual
shear stress acting on the membrane when there are fluids of
differing viscosities on either side of it. In this case, the
shear stress associated with the compression mode will be
proportional to the sum of the viscosities on either side of the
membrane, while the shear stress associated with the bending mode
will be proportional to the difference in those viscosities.
Taking these observations into account, we find that the shear
stress from the fluid at the surface of the membrane may be
written as
\begin{equation}\label{shear-answer}
\left. \sigma^{\rm f}_{\alpha z} \right|_{z=0} = 2 \omega q_\alpha
\left[ i \left( \Sigma \eta \right) u^{\rm L}(q) - \frac{1}{2}
\left( \Delta \eta \right) h(q) \right].
\end{equation}

In a similar way we may compute the normal stresses on the
membrane due to the hydrodynamic flows. The stress component
normal to the membrane is
\begin{equation}\label{normal-fluid}
  \sigma^{\rm f}_{zz} = 2 \eta \partial_z v_z - P
\end{equation}
where $P$ is the hydrostatic pressure computed from the fluid
velocity field and the Stokes equation.  Here we find that only
the bending mode generates a nonvanishing normal stress component.
The compression mode produces a pressure variation across the
surface of the membrane that exactly cancels the $z$-derivative of
the vertical velocity field.  We write the normal stress as
\begin{equation}\label{normal-stress}
  \left. \sigma^{\rm f}_{zz} \right|_{z=0} = 2 i \omega \left(
  \Sigma \eta \right) |q| h(q).
\end{equation}

Returning to the Fourier--transformed force balance equation we
write out the components in the $\hat{q}$ $\hat{z}$ subspace.  The
$\hat{z}$ equation is identical to Eq.~(\ref{bending-amplitude}).
We use this result to eliminate the dependence in the $\hat{q}$
equation, which takes the form:
\begin{eqnarray}\label{qhat}
  B q^2 u^{\rm L}(q) - && 2 q \omega \left[ i {\rm
  sign} (q) \left( \Sigma \eta \right) u^{\rm L}(q) - \frac{1}{2}
  \left(\Delta \eta \right) h(q) \right]\nonumber \\
  && = {\bf f} (q) \cdot \hat{q}
\end{eqnarray}
where ${\bf f} (q)$ is the externally applied force and $B = 2 \mu
+ \lambda$. Solving for the longitudinal part of the displacement
field we arrive at
\begin{eqnarray}\label{u-long}
 u^{\rm L}(q)&=& \frac{ {\bf f} (q) \cdot \hat{q}}{B q^2 - 2 i
 \omega |q| \left( \Sigma \eta \right) }\\ \nonumber
 & &  - \frac{\omega |q|
 \left(\Delta \eta \right) f_z(q)}{\left[ \kappa q^4 - 2 i \omega  \left( \Sigma \eta \right)
 \right] \left[B q^2 - 2 i
 \omega |q| \left( \Sigma \eta \right) \right]}.
\end{eqnarray}

Combining the above result for the longitudinal part of the
in--plane displacement field with the previously calculated
transverse part of the in--plane displacement field as well as the
out--of--plane perpendicular displacement, we can write the
trajectory of any point on the membrane as:
\begin{equation}\label{trajectory}
  {\bf R}(x_\alpha, \omega) = \hat{z} h(x_\alpha,\omega) + {\bf
  u}(x_\alpha, \omega),
\end{equation}
which in the Fourier transformed variable takes the form
\begin{equation}\label{trajectory-q}
R_i(q_\alpha, \omega) = \hat{z}_i h(q_\alpha,\omega) + \hat{q}_i
u^{\rm L}(q_\alpha,\omega) + \left( \delta_{i \beta} - \hat{q}_i
\hat{q}_\beta \right) u_\beta(q_\alpha ,\omega).
\end{equation}

We now calculate the one- and two--particle response functions for
the asymmetric membrane using the above results.  To do this, we
localize the applied force at the origin by taking ${\bf
f}(x_\alpha,\omega) = \exp(- i \omega t) {\bf f} \delta({\bf x})$
and we calculate the displacement field of a particle attached the
membrane at some other location ${\bf {\cal X}}$ using the
relation
\begin{equation}\label{observed-point}
  {\bf R}({\bf {\cal X}}, \omega) = \int \frac{d^2q}{(2 \pi)^2} {\bf R}(q_\alpha,\omega)
  e^{i {\bf q} \cdot {\bf {\cal X}}}.
\end{equation}
From this calculation we determine the response function tensor
$\alpha_{ij}({\bf {\cal X}},\omega) $ defined by the equation
\begin{equation}\label{response-tensor}
  R_i({\bf {\cal X}},\omega ) = \alpha_{ij}({\bf {\cal X}},\omega)
  f_j({\bf 0},\omega),
\end{equation}
so that $\alpha_{ij}({\bf {\cal X}},\omega) $ measures the
displacement of a point at ${\bf {\cal X}}$ on the undeformed
membrane in response to a force applied at the origin of the
coordinate system.  As before, we take into account the finite
size of the particle at the origin (where the force is applied)
and the tracer particle at ${\bf {\cal X}}$ by cutting off the
wavevector integrals at a maximum wavevector proportional to the
inverse particle radius.  We choose the coefficient of
proportionality to be the same as we have already used above.  For
concreteness we take the displacement vector of the observation
point at ${\bf {\cal X}}$ to be along the $\hat{x}$--axis.  It is,
of course, trivial to rewrite the resulting expressions in an
arbitrary reference frame.

Using Eqs.~(\ref{u-long}),(\ref{trajectory-q}), our previous
solutions for the bending and transverse, in--plane membrane
deformations in Eq.~(\ref{observed-point}) we first determine the
displacement of the tracer particle in  $\hat{x}$ direction.  The
resulting angular integrals are simply written as Bessel functions
and we find
\begin{equation}\label{Rx}
  R_x({\cal X},\omega) = \alpha_{xx}({\cal X},\omega) f_x + \alpha_{xz}({\cal
  X},\omega)f_z
\end{equation}
where the components of the response tensor are given by the
integrals over the magnitude of the wavevector $q$
\begin{eqnarray}\label{axx}
\alpha_{xx}({\cal X},\omega) &=& \frac{1}{4 \pi} \int_0^{c/a} d q
\, q \left\{ \frac{J_0(q |{\cal X}|) - J_2(q |{\cal X}|)}{B q^2 -
2 i \omega |q| (\Sigma \eta ) } + \right.  \\ \nonumber
 &  &\left.  + \frac{J_0(q |{\cal
X}|) + J_2(q |{\cal X}|)}{\mu q^2 -  i \omega |q| (\Sigma \eta ) }
\right\}\\
\label{axz}
\alpha_{xz}({\cal X},\omega) &=& \frac{-i
\omega (\Delta \eta)}{2 \pi} \int_0^{c/a} d q \, q^2 \frac{J_1(q
|{\cal X}|)}{\left[B q^2 - 2 i \omega |q| (\Sigma \eta ) \right] }
\nonumber \\ & & \times \frac{1}{\left[\kappa q^4 - 2 i \omega |q|
(\Sigma \eta ) \right]}.
\end{eqnarray}
The most interesting point of this result is the presence of a
finite $\alpha_{xz}$ in this problem.  We see that, due to the
hydrodynamic coupling through the subphase/superphase, a force
applied normal to the membrane at the origin produces in--plane
displacements at a different point ${\cal X}$.  This in--plane
motion is due to the coupling of membrane compression modes to the
bending of the membrane generated by the applied force in the
$\hat{z}$--direction.  The appearance of this coupling is clearly
proportional to the viscosity difference between the subphase and
the superphase and vanishes in the symmetric membrane problem.
This hydrodynamic coupling is only observed in the two--point
measurement; in the limiting case where one observes the
displacement of the particle at the origin due to a force applied
to it (the $|{\cal X}| \longrightarrow 0$ limit of the two--point
measurement which is clearly identical to the one--point
measurement in this calculation) the off--diagonal part of the
response function vanishes since $J_1(0) = 0$.  The vanishing of
the off-diagonal component is clearly required by the azimuthal
symmetry about the $\hat{z}$--axis of the single particle problem.

The remaining nonvanishing components of the response function
tensor are easily calculated in an analogous manner.  We find that
these components are given by
\begin{eqnarray}\label{ayy}
\alpha_{yy}({\cal X},\omega) &=& \frac{1}{4 \pi} \int_0^{c/a} d q
\, q \left\{ \frac{J_0(q |{\cal X}|) + J_2(q |{\cal X}|)}{B q^2 -
2 i \omega |q| (\Sigma \eta ) } + \right.  \\ \nonumber
 &  &\left.  + \frac{J_0(q |{\cal
X}|) - J_2(q |{\cal X}|)}{\mu q^2 -  i \omega |q| (\Sigma \eta ) }
\right\}
\end{eqnarray}
for in--plane motion due to an in--plane force perpendicular to
the separation vector of the two particles.  For displacements
normal to the plane of the membrane we find
\begin{equation}\label{azz}
\alpha_{zz}({\cal X},\omega) = \frac{1}{2 \pi} \int_0^{c/a} d q \,
q \frac{J_0(q |{\cal X}|)}{\kappa q^4 - 2 i \omega (\Sigma \eta )
|q|}.
\end{equation}

It should be noted that for large values of their arguments,
$J_0(x) + J_2(x)$ is dominated by $J_0(x) - J_2(x)$, so the
response function along the line of centers between the two
particles is dominated by the compression modulus while the
response function for in--plane motion perpendicular to the line
of centers is controlled by the shear modulus as one would expect.
In the next section we turn to the fluctuation--dissipation
theorem to compute the experimentally observable correlated
thermal fluctuations of two tracer particles embedded in the
membrane.

\section{Position correlation spectra}\label{experimental}

Having computed the response function to an applied force for both
single particle and two--particle systems, we now turn to the
question of the what is the experimentally accessible quantity.
There are two basic types of microrheological measurements that
are possible. In active microrheology in which the response
function is directly probes via the linear response measurement of
the displacement of a tracer due to a force applied either to that
particle (one particle measurements) or to another particle
embedded in the membrane (two particle measurements).  The
predicted response functions measured in these experiments have
been directly calculated in this paper.  It is, however, also
possible to use the correlated thermal fluctuations of a single
particle or two particles embedded in the membrane to access the
same rheological information via the Fluctuation--Dissipation
theorem.  In this section we concentrate on the correlated motion
of two particles embedded in the membrane in order to highlight
the effect of the hydrodynamic coupling between the bending and
compression modes that, we believe constitute a novel effect. The
quantity that we compute is
\begin{equation}\label{correlation}
  S_{ij}({\cal X},\omega) = \int  \langle R_i({\cal X},t)
  R_j({\bf 0},0) \rangle e^{i \omega t} d \, t.
\end{equation}

By the Fluctuation--Dissipation theorem this correlation function
can be written in terms of the imaginary part of the response
functions calculated above as
\begin{equation}\label{FDT}
S_{ij}({\cal X},\omega) = \frac{2 k_{\rm B} T}{\omega} \alpha''_{i
j}({\cal X},\omega),
\end{equation}
where $\alpha''_{i j}({\cal X},\omega)$ is the imaginary part of
the response (in the $i^{\rm th}$ direction) of a particle
embedded in the membrane at ${\cal X}$ to a force in the $j^{\rm
th}$ direction. It only remains for us to compute the remaining
integrals over the magnitude of the wavevector $q$ to determine
the correlation spectra.

For concreteness we put the vector defining the separation of the
particles along the $\hat{x}$ direction.  We first compute the
correlations of the in--plane motion of the particles
perpendicular to their line of centers and along their line of
centers by calculating $S_{yy}({\cal X},\omega)$ and $S_{xx}({\cal
X},\omega)$ respectively. In all of the following calculations we
assume that the in--plane shear modulus is dominated by the real
part and that this is frequency--independent.  In other words we
consider the membrane to act like a perfectly elastic sheet. Other
calculations can, of course, be performed with the formulae given
above.  We present these results merely as an example of the
correlation functions for this particularly simple case.  More
complex assumptions about the viscoelasticity of the membrane,
presumably based on microscopic models of the membrane, can of
course be incorporated in the formalism provided and the resulting
integrals can be performed.

We may write the imaginary part of the response function in the
following  form for the motion perpendicular to the line of
centers as
\begin{equation}\label{ayy-imaginary}
  \alpha''_{yy} = \frac{1}{4 \pi \mu \tau} \int_0^{c |{\cal X}|/a} d
  \, z \left\{ \frac{J^-(z)}{z^2 + \tau^{-2}} + \left( \frac{\mu}{B}
  \right)^2 \frac{2 J^+(z)}{z^2 + {\tau'}^{-2}} \right\}
\end{equation}
where we have defined the functions $J^{\pm}(z)$ to be the
following combinations of Bessel functions of the first kind:
\begin{equation}\label{J+-}
  J^\pm(z) = J_0(z) \pm J_2(z)
\end{equation}
and $\tau$ and $\tau'$ are a frequency--dependent functions of the
form
\begin{eqnarray}\label{tau-def}
  \tau^{-1} &=& \frac{\omega (\Sigma \eta) |{\cal X}|}{\mu}\\
  \label{tau-prime-def}
{\tau'}^{-1} &=& \frac{2 \omega (\Sigma \eta) |{\cal X}|}{B}.
\end{eqnarray}
Using the same definitions, we write the imaginary part of the
response function for motion along the line of centers as
\begin{equation}\label{axx-imaginary}
  \alpha''_{xx} = \frac{1}{4 \pi \mu \tau } \int_0^{c |{\cal X}|/a} d
  \, z \left\{ \frac{J^+(z)}{z^2 + \tau^{-2}} + \left( \frac{\mu}{B}
  \right)^2 \frac{2 J^-(z)}{z^2 + {\tau'}^{-2}} \right\}
\end{equation}

The physical interpretation of $\tau$ is clear: it measures the
ratio of the distance separating the two particles to the
screening length of shear modes in the membrane due to the
coupling to the viscous subphase and superphase.  The other
function, $\tau'$, has an analogous meaning in terms of the
compression modes of the membrane.  Up to an overall scale factor
the shape of the correlation spectrum obeys a distance--frequency
scaling relation so that spectra obtained at different
interparticle separations can be collapsed unto a master curve by
rescaling distances so they are measured in terms of the
frequency--dependent screening length.  The breakdown of this
scaling relation can be used as a diagnostic of the appearance
some significant frequency--dependence of the membrane elastic
moduli.

The only distinction between the two response functions shown
above is interchange of the roles of the functions $J^\pm(z)$. The
effect of this switch is that the compression response dominates
the long--length scale response along the line of centers while
the shear response of the membrane controls the long--length scale
response perpendicular to the line of centers.  This observation
if intuitively obvious.  We plot the resulting correlation
spectra as a function of the dimensionless variable $\tau$ in the
limit that the compression modulus, $B = 2 \mu + \lambda$ is much
larger than the shear modulus in the membrane and that the
particle size is much smaller than the interparticle separation,
$|{\cal X}| \gg a$.  The plots of the undimensionalized
correlation functions: $I_{ij} = 4 \pi \mu^2 S_{ij}/(2 k_{\rm B}
T (\Sigma \eta) |{\cal X}|)$ are shown in figure \ref{Sxxyy-plot}.

\begin{figure}[bp]
\scalebox{0.3}{\includegraphics{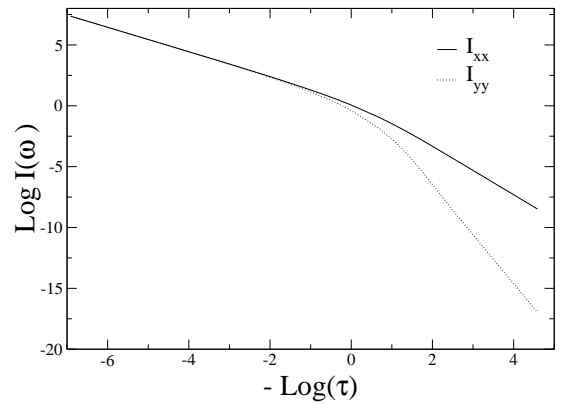}} \caption{The graph
of the dimensionless interparticle correlation function (see text)
for motion along the line of centers $I_{xx}$ and motion
perpendicular to the line of centers $I_{yy}$ as a function of the
dimensionless variable $\tau$ in the combined limits of small
particle size compared in the interparticle separation and high
membrane bulk modulus compared to the membrane shear modulus.}
\label{Sxxyy-plot}
\end{figure}

We now consider the same calculation for the correlated
out--of--plane fluctuations of the particles.  Once again using
the Fluctuation--Dissipation theorem with the appropriate
component of the response tensor, $\alpha_{zz}({\cal X},\omega)$,
\begin{equation}\label{azz-imaginary}
\alpha''_{zz} = \frac{1}{4 \pi \mu \tau} \int_0^{c |{\cal X}|/a} d
  \, z \frac{J_0(z)}{z^6 + {\tau''}^{-2}},
\end{equation}
where we have defined a new dimensionless, frequency--dependent
variable in analogy to $\tau$ and $\tau'$ above.  In this case,
$\tau''$ measures whether the viscous damping of the bending mode
on the length scale of the interparticle separation is relevant at
the frequency $\omega$.  It is defined by
\begin{equation}\label{taupp-def}
  {\tau''}^{-1} = \frac{2 \omega (\Sigma \eta ) |{\cal X}|^3}{\kappa}.
\end{equation}
We can write the undimensionalized correlation spectrum $I_{zz} =
\pi  \kappa^2 |{\cal X}|^6 S_{zz}({\cal X},\omega)/( 2 k_{\rm B} T
(\Sigma \eta) )$ in terms of an integral that is performed
numerically.  The resulting correlation spectrum is shown in
figure \ref{Szz-plot} where the independent variable is $\tau''$.
It may be noted the frequency, separation scaling property of the
fluctuations of the particles both along their line of centers and
perpendicular to their line of centers fails for the case of the
vertical motion.  This breakdown of the scaling reflects the
elementary result that bending energy is harmonic in the Laplacian
of the vertical displacement rather than in the gradient of the
displacement field as in the case of in--plane deformation.

\begin{figure}[bp]
\scalebox{0.3}{\includegraphics{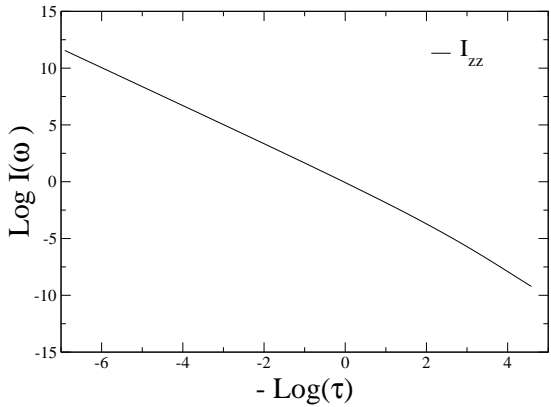}} \caption{The
graph of the dimensionless interparticle correlation function (see
text) for motion perpendicular to the membrane, $I_{zz}$ as a
function of the dimensionless variable $\tau''$ in the limit of
small particle size compared in the interparticle separation.}
\label{Szz-plot}
\end{figure}

Finally we consider the correlations between the out--of--plane
motion at one point and in--plane motion at another point.  These
correlations do not exist in the symmetric membrane which has a
vanishing $\alpha_{xz}$ component of the response tensor. However,
due to the existence of such a term in the asymmetric case, these
unexpected correlations will be present in the thermal
fluctuations of the compressible membrane.

Writing the imaginary part of the $\alpha_{xz}$ component of the
response tensor,
\begin{equation}\label{axz-imaginary}
\alpha''_{xz} = - \frac{ (\Delta \eta) \omega |{\cal X}|}{2 \pi
B^2} \int_0^{c |{\cal X}|/a} d
  \, z J_1(z) \frac{\kappa' z^4 - {\tau'}^{-2}}{\left(z^2 + {\tau'}^{-2} \right) \left( {\kappa'}^2 z^6 +
  {\tau'}^{-2}
  \right)},
\end{equation}
where we have used the definition of $\tau'$ from
Eq.~(\ref{tau-prime-def}) and have introduced the dimensionless,
interparticle--separation dependent quantity
\begin{equation}\label{kappa-prime-def}
  \kappa' = \frac{\kappa}{B |{\cal X}|^2},
\end{equation}
which measures the relative importance of the bending modulus and
the compression modulus at the length scale of the interparticle
separation $|{\cal X}|$.

\begin{figure}[bp]
\scalebox{0.3}{\includegraphics{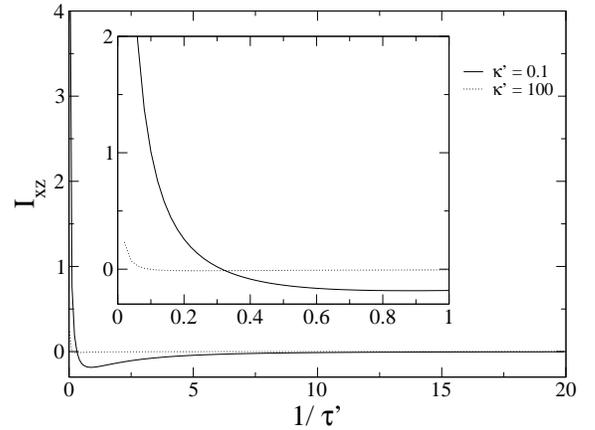}} \caption{The
graph of the dimensionless interparticle correlation function (see
text) for motion perpendicular to the membrane, $I_{xz}$ as a
function of the dimensionless variable $\tau''$ in the limit of
small particle size compared in the interparticle separation.}
\label{Sxz-plot}
\end{figure}

As we have done before, we plot the undimensionalized correlation
spectrum versus the dimensionless frequency variable, $\tau'$.  In
this case the dimensionless correlation variable takes the form:
$I_{xz} = - S_{xz}(|{\cal X}|,\omega)  \pi B^2/( k_{\rm B} T
(\Delta \eta ))$.  We see immediately that there are now a family
of curves generated by changing the dimensionless parameter
$\kappa'$. In figure \ref{Sxz-plot} we plot two representative
curves from this family: One at a low value of $\kappa'$
($10^{-1}$) and one at a high value of this parameter, $\kappa' =
10^2$.   In addition, due to the interparticle separation
dependence of this parameter, there is, once again, no
time--distance scaling as was found for the in--plane motion. This
correlation function is qualitatively different from the other
discussed above in that, at a fixed interparticle separation, it
changes sign as a function of frequency.  For concreteness we
momentarily assume that the superphase is less viscous than the
subphase, then, at low enough frequencies the vertical motion of
one particle is correlated with the displacement of the other one
along their line of centers. This is to be expected from the
hydrodynamic flow fields plotted earlier.  The upward motion of
one particle leads to an increase in the separation of the two
particles.  At higher frequencies, however, the size of the rolls
in the fluid changes so that the motion becomes anti-correlated.
Lastly, it is important to note that the overall magnitude of the
$x$ $z$ correlations depends on the in--plane bulk modulus. In
highly incompressible membranes, these correlations vanish as
$1/B^2$.

We now turn to one final power spectrum calculation.  Unlike the
ones discussed above, we now consider explicitly a viscoelastic
membrane. Based on the work of the Strasbourg group, we take as an
example, an actin--coated lipid membrane whose viscoelastic
properties are dominated by the actin coat.  We further specialize
to the high--frequency limit where the actin rheology is dominated
by single chain dynamics.  In this frequency range the effective
shear modulus of the membrane takes the form:
\begin{equation}\label{actin-G}
\mu(\omega) = \mu_{\rm a} (-i \hat{\omega})^{3/4}
\end{equation}
where $\mu_{\rm a}$ is a (real) modulus scale and $\hat{\omega} =
\omega \tau$ is a dimensionless frequency.  We now use
Fluctuation--Dissipation theorem in conjunction with the in--plane
response function given in Eq.~(\ref{answer-in-plane}) in which we
set the membrane shear modulus to that given above in
Eq.~(\ref{actin-G}) and the membrane compression modulus to
infinity.  This latter choice is based on the
near--incompressibility of the underlying lipid bilayer to which
the actin network is attached.

After some algebra we determine that the single particle power
spectrum is given by
\begin{equation}\label{actin-answer}
  \langle |x(\omega)|^2\rangle = \frac{ k_{\rm B} T}{2 \pi
  \mu_{\rm a} \tau^{3/4}} \omega^{-\frac{7}{4}} H(\omega)
\end{equation}
where the function $H(\omega)$ is given by:
\begin{eqnarray}\label{H}
  H(\omega) &=&  \cos \left( \frac{3 \pi}{8} \right) \arctan \left( \frac{\cos \left( \frac{3 \pi}{8} \right)
  \beta(\omega)}{1 + \sin \left( \frac{3 \pi}{8} \right)
  \beta(\omega)} \right) \\ \nonumber
    &   &  + \frac{1}{2} \sin \left( \frac{3 \pi}{8}
  \right)  \log \left[ 1 + \beta^2(\omega) + 2 \beta(\omega) \sin
  \left(\frac{3 \pi}{8} \right) \right].
\end{eqnarray}
The function $H(\omega)$ depends on frequency only through the
dimensionless quantity
\begin{equation}\label{beta-def}
  \beta(\omega) = \frac{2  \mu_{\rm a} \tau}{3 \eta a}
  \hat{\omega}^{-1/4}.
\end{equation}
In the limit of high frequency (small $\beta$) it is simple to
show that the power spectrum scales with frequency as
$\omega^{-2}$ as is expected for simple Brownian motion.  One can
show that in this limit the power spectrum takes the form
\begin{equation}\label{high-freq-limit }
  \langle |x^2(\omega)|\rangle \longrightarrow  \frac{ k_{\rm B} T}{3 \pi \eta a}
  \omega^{-2}, \, \mbox{as} \, \omega \rightarrow \infty.
\end{equation}

On the other hand, at intermediate frequencies high enough so that
Eq.~(\ref{actin-G}) accurately describes the membrane rheology,
but low enough so that $\beta$ is small, $H$ is independent of
frequency (up to logarithmic corrections) and we find that power
spectrum the tracer particle position fluctuations decays as a
different power law with frequency.  In this frequency range we
find that
\begin{equation}\label{intermediate-freq}
 \langle |x^2(\omega)|\rangle \sim \omega^{-7/4}.
\end{equation}

We also compute the power spectrum of out--of--plane fluctuations
for the membrane.  This result does not depend on the
frequency--dependent shear modulus of the actin.  Rather, it
depends on only  the bending modulus of the membrane, $\kappa$,
which, in principle, is also a complex, frequency--dependent
quantity.  In particular, for the actin--coated membrane, the
bending modulus takes into account the viscoelastic result of the
actin in addition to the usual bending energy of the lipid bilayer
so one should expect this quantity to have a complex frequency
response.  Inasmuch as this more microscopic mechanical issue has
not been satisfactorily  resolved, we will simply assume that the
bending modulus is a real, frequency--independent quantity.

The remaining calculation, of course, is done analogously to the
calculation presented above. We use Eq.~(\ref{out-of-plane-alpha})
and the Fluctuation--Dissipation theorem to compute the power
spectrum of $\hat{z}$--fluctuations. To do so we need to consider
the imaginary part of the response function by performing the
integral in Eq.~(\ref{out-of-plane-alpha}).  This integral depends
on the dimensionless parameter $\delta$ which is linearly
proportional to frequency.  In the low frequency limit, we find
that to leading order the imaginary part of integral takes the
form:
\begin{equation}\label{out-plane-integral}
{\rm Im}[\alpha_z(\omega)] = \frac{9 \pi a^2}{2 \kappa} \left[
\frac{\pi}{3}\delta^{-2/3} + {\cal O}\left( \delta^{-1/3} \right)
\right].
\end{equation}
From the Fluctuation--Dissipation theorem we then find that the
power spectrum in the low frequency limit takes the form:
\begin{equation}\label{low-freq-spectrum}
  \langle |z(\omega)|^2 \rangle = \frac{1}{3}  \left(\frac{2 \pi^2}{ \kappa \eta^2 }
  \right)^{1/3} k_{\rm B} T \omega^{-5/3}.
\end{equation}
Note that the above result is independent of tracer particle size.
The power--law decay of fluctuations with the above exponent has
already been calculated by Zilman and Granek\cite{Zilman:96}

\begin{figure}[bp]
\scalebox{0.32}{\includegraphics{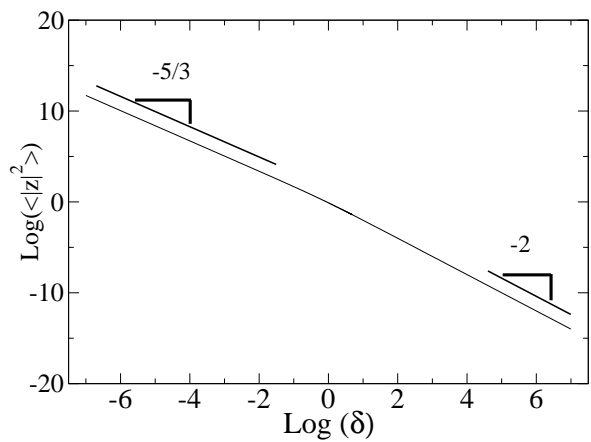}} \caption{The
out of plane tracer particle fluctuation spectrum in arbitrary
units plotted against a dimensionless frequency.  See the
definition of $\delta$ immediately following
Eq.~(\ref{out-of-plane-alpha}).  At low frequencies, where the
dynamics of the tracer is dominated by the membrane bending
stiffness, the $-5/3$ power law obtains.  At high frequencies
where the fluid viscosity dominates the dynamical response of the
particle , this power law crosses over to the $-2$ exponent, which
is expected for the Brownian motion of a free particle.}
\label{outofplanefig}
\end{figure}

In the high frequency limit, the effect on the tracer of the
membrane bending stress on the dynamics is dominated by the
viscous stress coming from the fluid.  The power spectrum for
Brownian fluctuations of the bead reduces to that of a {\it free}
Brownian sphere,
\begin{equation}\label{high-freq-spectrum}
\langle |z(\omega)|^2 \rangle \longrightarrow \frac{ k_{\rm B}
T}{3 \pi \eta a} \omega^{-2}, \, \, {\rm as} \, \, \omega
\longrightarrow \infty.
\end{equation}

The full power spectrum of the out--of--plane tracer particle
fluctuations is shown in figure~\ref{outofplanefig}; this plot
demonstrates the cross--over from the low--frequency, bending
stiffness dominated dynamics, to the high frequency dynamical
regime controlled by the viscous stresses in the surrounding
fluid.

\section{Summary}
\label{conclusions}

In this paper we have examined the dynamics of flat, viscoelastic
membranes either immersed a viscous, Newtonian fluid, or
separating two Newtonian fluids of differing viscosities.  We have
paid particular attention in this analysis to dynamical issues
related to microrheological measurements.  Thus, we have
calculated in some detail the correlated fluctuation spectrum of
two rigid particle embedded in the membrane.  In addition we have
computed the response  function of a single particle in such a
membrane. Both of these calculations can be directly applied to
analysis of experimental data, whether it be from a actual linear
response measurement of the force--position response of a tracer
particle ({\it e.g.\/} with magnetic particles, or laser tweezers)
or from the measurement of the autocorrelations or interparticle
correlations of tracer particles undergoing thermal, Brownian
motion on the membrane.

The range of application systems is quite large.  At the moment we
only  restrict the equilibrium shape of the viscoelastic membrane
to be flat, although we expect the results in this paper to be
applicable to curved systems as long as  the particle motion and
interparticle separation in the case of the two--particle
measurements remains small compared to the radius of curvature.
The above work can be applied to lamellar phases in aqueous
surfactant systems, lipid bilayers, emulsion droplets,
``polymersomes''(as long as they are large enough -- see the above
comments regarding curvature), Langmuir monolayers, and cell
membranes.

One of the principle results of this work is the prediction of a
novel cross--correlation between the vertical motion of one
particle and the in--plane motion of another one embedded in the
same membrane.  Such coupling between the out--of--plane motion of
one and the in--plane motion of the other particle is mediated
purely by the hydrodynamic interaction through the surrounding
liquid and vanishes (by symmetry) if the Newtonian fluids above
and below the membrane are identical.  Such cross--correlations
should be observable in a large variety of systems since in a
number of general multiphase and in particular  biological systems
membranes separate fluids of different viscosities. This effect,
for example, should be present in studies of cell membranes.  It
should be most dramatic, however, in cases of high membrane
compressibility and large fluid viscosity differences across the
membrane.  The best system then to look for such
cross--correlations will be a compressibility Langmuir monolayer.

A number of extensions of this work are both possible and
interesting to pursue.  The first is to study the role of
equilibrium membrane curvature on the dynamics.  Since it is
already known that there is a purely geometric coupling between
in--plane strains and out--of--plane bending modes, one should
find a rich structure in the cross--correlations of in--plane and
out--of--plane motion resulting from the interplay of the
geometric and hydrodynamic coupling of these modes.  In addition,
such an analysis is necessary to extend the methods of
microrheology to highly curved surfaces.  Other important
extensions of the present theory include the analysis of the
coupling of a viscoelastic membrane to a viscoelastic bulk
material.  The cell membrane coupled to the viscoelastic
cyctoplasm is an important realization of such a system.

Along the lines of applying these results to cellular
microrheology, we have to acknowledge that we have studied a
one--component, continuum model of the membrane, whereas the
physical cell membrane is a highly heterogeneous material on the
submicron length scale.  It is therefore important to examine a
more complex, multi-component model for the membrane.  Such more
detailed dynamical models have been previously studied in the
context of three--dimensional microrheology. These studies in
according with very general arguments have shown that the there
are necessarily extra dynamical modes in the heterogeneous models.
However, based on the three--dimensional work, we expect that the
principle effect of these extra modes will be to change the
correlation spectra only at the lowest measurable frequencies.

We thank D.~Pine, R.~Granek, J.F.~Joanny, D.~Lubensky,
T.C.~Lubensky, E.~Helfer, C.F.~Schmidt, L.~Bourdieu, D.~Weitz,
D.~Wirtz and D.~Chatenay for useful discussions. AJL is
particularly grateful for the hospitality of the Vrije
Universiteit Division of Physics and Astronomy where much of this
work was done. This work was supported in part by the National
Science Foundation under Award DMR-9870785.

\end{document}